\documentclass[prl,aps,showpacs,twocolumn,superscriptaddress,floatfix]{revtex4-1}

\usepackage{epsfig}
\usepackage{amsmath}
\usepackage{amssymb}
\usepackage{amsfonts}
\usepackage{mathptmx}
\usepackage{eucal}
\usepackage{bm}
\usepackage{color}

\usepackage{epstopdf}

\usepackage{graphicx}

\usepackage{natbib}

\def\rv{{\bf r}}

\def\nv{{\bf n}}
\def\pv{{\bf p}}

\def\Rv{{\bf R}}

\def\Fcal{{\mathcal F}}
\def\Acal{{\mathcal A}}
\def\tr{{\rm tr}}

\newcommand{\ctG}{{\mathcal{\check{\tilde{G}}}}}

\def\tf{{\tilde{f}}}

\newcommand{\cG}{{\mathcal{\check G}}}

\newcommand{\sgn}{{\mathrm{sgn}}}

\begin{document}
\title{Singlet-triplet conversion and the long-range proximity effect in superconductor-ferromagnet structures with generic spin dependent fields}
\author{F. S. Bergeret}
\affiliation{Centro de F\'{i}sica de Materiales (CFM-MPC), Centro
Mixto CSIC-UPV/EHU, Manuel de Lardizabal 4, E-20018 San
Sebasti\'{a}n, Spain}
\affiliation{Donostia International Physics Center (DIPC), Manuel
de Lardizabal 5, E-20018 San Sebasti\'{a}n, Spain}
\affiliation{Institut f\"ur Physik, Carl von Ossietzky Universit\"at, D-26111 Oldenburg, Germany}
\author{I. V. Tokatly}
\affiliation{Nano-Bio Spectroscopy group, Dpto. F\'isica de Materiales, Universidad del Pa\'is Vasco, Av. Tolosa 72, E-20018 San Sebasti\'an, Spain}
\affiliation{IKERBASQUE, Basque Foundation for Science, E-48011 Bilbao, Spain}

\date{\today }

\begin{abstract}
The long-range proximity effect in superconductor/ferromagnet (S/F) hybrid nano-structures is observed if singlet Cooper pairs from the superconductor are converted into triplet pairs which can diffuse into the ferromagnet over large distances. It is commonly believed that this happens only in the presence of magnetic inhomogeneities. We show that there are other sources of the long-range triplet  component (LRTC) of the condensate and establish general conditions for their occurrence. As a prototypical example we consider  first a system where the exchange field and spin-orbit coupling can be treated as time and space components of an effective SU(2) potential. We derive a SU(2) covariant diffusive equation for the condensate and demonstrate that an effective SU(2) electric field is responsible for the long-range proximity effect. Finally,  we extend our analysis to a generic ferromagnet and establish a universal condition for the LRTC. Our results open a new avenue in the search for such correlations in S/
F structures and make a hitherto unknown connection between the LRTC and Yang-Mills electrostatics.
\end{abstract}
 
\maketitle

\bigskip

The odd triplet superconductivity in superconductor-ferromagnet (S/F) structures has been intensively studied, both theoretical and experimentally, since its prediction in 2001 \cite{Bergeret2001,BergeretRMP}.  
In that context it is of crucial interest to understand the  process of converting the singlet Coppers pairs from the superconductor into triplet pairs of electrons with equal spins in the ferromagnet.
Apart from its interest for fundamental research, the study of triplet superconducting correlations might find useful applications for spintronics \cite{EschrigPT}.
 
It is well established that  triplet pairs, once created,  diffuse into the ferromagnetic materials over  distances much larger than the singlet ones. 
This  leads to the  long-range proximity effect which explains  the observation of  Josephson currents through  S/F/S junctions over large distances \cite{Keizer2006,Birge2010,Robinson2010,Aarts2010, Wang2010,Sprungmann2010,Birge2012} and the long-range propagation in S/F structures of  superconducting correlations  \cite{Villegas2012} . The usual theoretical interpretation of these experiments assumes that the singlet-triplet conversion is mediated by a magnetic inhomogeneity in the vicinity of the  S/F  interface\cite{inhomo}. This can be caused by a domain wall \cite{Bergeret2001}, a spin-active S/F  interface\cite{Eschrig2008}, or by a multilayered ferromagnetic structure with different magnetic orientations \cite{Bergeret2003}.

Formally, the existence of the long-range triplet component (LRTC) can be inferred by inspecting the spin structure of the  quasiclassical condensate (anomalous) Green's function (GF), $\hat f=f_s+ {\bf f}_t\hat{\bf\sigma}$.
Here $f_s$ is the amplitude of the  singlet component, and the vector ${\bf f}_t$ describes the  triplet correlations  ($\hat{\bf \sigma}$ is the vector of Pauli matrices). The component of the vector ${\bf f}_t$ parallel to the magnetization describes the triplet state with zero spin-projection (i.~e., $|\uparrow,\downarrow\rangle + |\downarrow,\uparrow\rangle$)\cite{Champel2005,Fominov2006}.  The LRTC corresponding to the pairs with 
spin-projections $\pm 1$ only exists if ${\bf f}_t$ is non-collinear with the magnetization direction, which is  the case for certain magnetic inhomogeneities \cite{Champel2005,Fominov2006,Note0}. Indeed, it is commonly believed that the only way to generate the LRTC in S/F hybrids is by means of creating  a non-homogeneous magnetic configuration.   

In this Letter we study  the  LRTC in S/F structures from a more general perspective and demonstrate that, besides  an inhomogeneous magnetization, there are other sources of long-range triplet correlations. In particular, the momentum dependence of an effective exchange field,which can be attributed to the spin-orbit (SO) coupling naturally generates LRTC, provided certain conditions are fulfilled.
We also show that the  physical mechanism of the singlet-triplet conversion can be linked to the local SU(2) invariance of magnetized systems with SO interaction \cite{Mineev1992,Frolich1993}. To reveal this physics we first analyze a prototypical example of a spin-dependent field consisting of a momentum independent Zeeman term and a linear in momentum  SO coupling. These two contributions act as the time and space components of the SU(2) gauge potential, which ensures that that the Zeeman and SO fields  enter physical quantities only in gauge covariant combinations \cite{Tokatly2008}. We derive a SU(2) covariant Usadel equation and identify the SU(2) electric field as a key object responsible for the long-range triplet proximity effect and for the existence of the LRTC in S/F structures. By solving the Usadel equation for a lateral S/F junction we demonstrate that even in the case of a uniform exchange field, the LRTC is present in the system. 
In the second part, we generalize our results to systems with possibly anisotropic Fermi surfaces and generic spin-dependent  fields with an arbitrary momentum dependence.  We derive the general quasiclassical equations for the anomalous GF, which allows us to establish  universal conditions for the creation of triplet long-range superconductivity in diffusive S-F hybrids.  

  We consider a hybrid structure consisting of a conventional s-wave superconductor (S) with the order  parameter $\Delta$ {in contact with a ferromagnet (F) with a SO coupling}. Let us first assume that SO effects can be modeled by a generic linear in momentum form $H_{SO}=\frac{1}{2m}\{p_j,\hat{\Acal}_j\}$, where $\hat{\Acal}_j=\Acal_j^a\sigma^a$ are the components of a $2\times 2$ matrix valued vector that parametrizes the SO coupling. In this case the Hamiltonian in the F-region can be represented in the form
\begin{equation}
\label{H-linearSO}
H = \frac{1}{2m}(p_j - \hat{\Acal}_j)^2 - \hat{\Acal}_0+V_{imp}, 
\end{equation}
where $\hat{\Acal}_0=\Acal_{0}^a\sigma^a\equiv h^a\sigma^a$ is the exchange field, and $V_{imp}$ is the {spin-independent} impurity scattering term \cite{noteSO}. The SO coupling and the Zeeman term enter the problem as the space and time components of the SU(2) gauge potential, which implies the SU(2) gauge invariance. The Hamiltonian remains unchanged under any local SU(2) rotation with a matrix $\hat{U}(\rv)$ supplemented with and the gauge transformation of the potentials,
$ \hat{\Acal}_j\mapsto\hat{U}\hat{\Acal}_j\hat{U}^{-1} -i (\partial_j\hat{U})\hat{U}^{-1}$,  $ \hat{\Acal}_0\mapsto\hat{U}\hat{\Acal}_0\hat{U}^{-1}$.
Since we are interested in equilibrium quantities we work with the Matsubara 4$\times$4 matrix (in the Nambu$\times$spin space) GF $\cG_\omega(\rv_1,\rv_2)$ at the discrete frequencies $\omega=\pi T(2n+1)$. 
To keep track of exact SU(2) gauge symmetry we employ a technique developed in the context of quark-gluon kinetics \cite{Heinz} and used recently to describe spin dynamics in semiconductors \cite{Gorini2010}.
Namely, we introduce the covariant GF as follows
$ \ctG_\omega(\rv_1,\rv_2) = \hat{W}(\Rv,\rv_1)\cG_\omega(\rv_1,\rv_2)\hat{W}(\rv_2,\Rv)$
where $\hat{W}(\Rv,\rv_1)$ and $\hat{W}(\rv_2,\Rv)$ are the Wilson link operators which ``covariantly connect'' the arguments of the GF to the "center-of-mass" coordinate  $\Rv=\frac{\rv_1+\rv_2}{2}$ \cite{Note}. The advantage of the covariant GF is that its Wigner transform,  and thus the corresponding quasiclassical GF $\check{\tilde g}(\nv,\Rv)$,  transform locally covariantly under a nonuniform SU(2) rotation, i.~e. $\check{\tilde g}\mapsto\hat{U}\check{\tilde g}\hat{U}^{-1}$. By using the method of Ref.\cite{Heinz} we can derive the equation of motion for $\ctG_\omega$ and then proceed further to the quasiclassical limit and eventually to the diffusive Usadel equation \cite{Usadel}. Here we only show the final linearized Usadel equation {in the ferromagnet} for the covariant anomalous function $\hat{\tf}$
\begin{equation}
 \label{covUsadel-gen}
D\tilde{\nabla}_{k}(\tilde{\nabla}_{k}\hat{\tf}) - 2|\omega|\hat{\tf} -i\sgn\omega \{\hat{\Acal}_0,\hat{\tf}\}=0\; .
\end{equation}
Here  $\tilde{\nabla}_{k}$  is the covariant gradient operator defined by $\tilde{\nabla}_{k}\Psi=\partial_{k}\Psi -i [\hat{\Acal}_k,\Psi]$ . {At the S/F boundary   we use the  Kupriyanov-Lukichev  boundary conditions \cite{KL} which  take the form}
\begin{equation}
 \label{covBC-gen}
N_{k}\tilde{\nabla}_{k}\hat{\tf}\big\vert_I = -\gamma f_{\Delta},
\end{equation}
where $f_{\Delta}=\Delta/\sqrt{\omega^2+\Delta^2}$ is the anomalous GF in the S-region and $N_k$ the $k$-component of the  vector normal to the S/F interface.  Hence in the covariant formalism the usual gradients are replaces by the covariant ones, and this is the only place where the SO coupling enters the theory. { Eqs.~(\ref{covUsadel-gen}), (\ref{covBC-gen}) are manifestly gauge covariant and
 their structure  is very appealing physically.  In fact, they can be written immediately by using only the gauge symmetry arguments \cite{Note1}.} 

Next, we write $\hat{\tf}$ as the sum of the singlet $f_s$ and triplet $\hat{\tf}_t$ contributions, splitting out  of $\hat{\tf}_t$ the part parallel to the exchange field $\hat{\Acal}_0$:
\begin{equation}
 \label{f-split}
 \hat{\tf} = f_s + \hat{\tf}_t = f_s + \hat{\Acal}_0 f_t^{\parallel} + \hat{\tf}_t^{\perp}\; .
\end{equation}
For any matrix $\hat{M}$ we have defined $\hat M^{\perp}= \frac{1}{4}[\hat{\Acal}_0,[\hat{\Acal}_0,\hat{M}]]/|\hat{\Acal}_0|^2$, where $|\hat{\Acal}_0|=\sqrt{\Acal^a\Acal^a}$ is the amplitude of the exchange field.

Trace of Eqs.~(\ref{covUsadel-gen}), (\ref{covBC-gen}) gives the equations for the 
singlet amplitude $f_s$ coupled to the parallel triplet amplitude $f_t^{\parallel}$:
\begin{equation}
  \label{covUsadel-s}
\nabla^2f_s - \kappa_\omega^2f_s -2i\frac{\sgn\omega}{D} |\hat{\Acal}_0|^2f_t^{\parallel}=0,\quad
N_{k}\partial_{k}f_s\big\vert_I = -\gamma f_{\Delta}
\end{equation}
where $\nabla^2$ is the usual Laplace operator, and $\kappa_\omega^2=2|\omega|/D$. The traceless part of Eqs.~(\ref{covUsadel-gen}), (\ref{covBC-gen}), can be rearranged as follows
\begin{eqnarray}
 \nonumber
&&\hat{\Acal}_0\left[\nabla^2f_t^{\parallel} - \kappa_\omega^2 f_t^{\parallel} -i2\frac{\sgn\omega}{D}f_s\right]\\
 \label{covUsadel-t}
&+&\left[\tilde{\nabla}_{k}(\tilde{\nabla}_{k}\hat{\tf}_t^{\perp}) - \kappa_\omega^2\hat{\tf}_t^{\perp}
+ 2\hat{\Fcal}_{k0}\partial_{k}f_t^{\parallel} + f_t^{\parallel}\tilde{\nabla}_{k}\hat{\Fcal}_{k0}\right]=0,\\
\label{covBC-t}
&&N_{k}(\hat{\Acal}_0\partial_{k}f_t^{\parallel} + \tilde{\nabla}Ì£_{k}\hat{\tf}_t^{\perp}
+ \hat{\Fcal}_{k0}f_t^{\parallel})\big\vert_I = 0.
\end{eqnarray}
Here  $\hat{\Fcal}_{k0}$ is the SU(2) electric field defined as
\begin{equation}
\hat{\Fcal}_{k0} = \partial_{k}\hat{\Acal}_0 - i [\hat{\Acal}_k,\hat{\Acal}_0].
\label{su2e}
\end{equation}
Inspection of Eqs. (\ref{covUsadel-t}-\ref{covBC-t}) shows that $\hat{\Fcal}_{k0}$  is the key object for the long range proximity effect: If  $\hat{\Fcal}_{k0}=0$  the homogeneity of the boundary condition (\ref{covBC-t}) implies that $\hat{\tf}_t^{\perp}=0$. In other words, no transverse triplet is generated in the absence of the SU(2) electric field. {$\hat{\Fcal}_{k0}$ vanishes in the case of a uniform magnetization and if the  SO and Zeeman terms commute.}

If $\hat{\Fcal}_{k0}$ is small Eqs.~(\ref{covUsadel-t}) and (\ref{covBC-t}) can be treated perturbatively. The leading contribution is given by the terms proportional to $\hat{\Acal}_0$, i.~e., by the first line in Eq.~(\ref{covUsadel-t}) and by the first term in Eq.~(\ref{covBC-t}). This yields the well known  equations for the short-range triplet component $f_t^{\parallel}$ coupled to the singlet  one $f_s$,
\begin{equation}
 \label{covUsadel-tpar}
\nabla^2f_t^{\parallel} - \kappa_\omega^2f_t^{\parallel} -2i\frac{\sgn\omega}{D} f_s=0,\quad
N_{k}\partial_{k}f_t^{\parallel}\big\vert_I = 0.
\end{equation}
Equations for the lowest in $\hat{\Fcal}_{k0}$ correction to $\hat{\tf}_t$ come from the second line in Eq.~(\ref{covUsadel-t}) and the last two terms in Eq.~(\ref{covBC-t}):
\begin{eqnarray}
 \label{covUsadel-tperp}
&&\nabla^2\hat{\tf}_t^{\perp} - \kappa_\omega^2\hat{\tf}_t^{\perp} =
- \left[ 2\hat{\Fcal}_{k0}\partial_{k}f_t^{\parallel} + f_t^{\parallel}\tilde{\nabla}_{k}\hat{\Fcal}_{k0}\right]^\perp,\\
\label{covBC-tperp}
&&N_{k}(\partial_{k}\hat{\tf}_t^{\perp} + \hat{\Fcal}_{k0}^\perp f_t^{\parallel})\big\vert_I = 0.
\end{eqnarray}
In the first terms in the left hand sides we replaced $\tilde{\nabla}_{k}$ with $\partial_k$ as the difference of these operators gives higher order corrections compared to those determined by the terms $\sim f_t^{\parallel}$.

Equations (\ref{covUsadel-s}), (\ref{covUsadel-tpar})-(\ref{covBC-tperp}) provide a complete description of S/F structures, which clearly demonstrates a common physical role of the SO coupling and inhomogeneous magnetization in the problem of the singlet-triplet conversion. As expected on general grounds, they appear in the theory in form of a single gauge covariant object -- the SU(2) electric field $\hat{\Fcal}_{k0}$ Eq.~(\ref{su2e}) entering the ``source part'' of Eq.~(\ref{covUsadel-tperp}) for $\hat{\tf}_t^{\perp}$. 

It is easy to see that the well known generation of LRTC by magnetic inhomogeneities follows naturally from our covariant formulation. Consider for example a transversal multilayer S/F structure shown in Fig~1a in the absence of SO coupling. In this case a nonzero $\hat{\Fcal}_{k0}$ is solely due to inhomogeneity of the exchange field $\hat{\Acal}_0$. {Assume that $\hat{\Acal}_0$ has only an  in-plane component (eventually rotating).} Then the first term in the r.h.s. of Eq. (\ref{su2e}) generates the LRTC for a Bloch domain wall parallel to the interface \cite{Bergeret2001}, while the second term $\sim\tilde{\nabla}_{k}\hat{\Fcal}_{k0}$ is responsible for the LRTC in the presence of a finite Neel wall perpendicular to the interface plane \cite{Fominov2005}.  It is interesting to note that the covariant derivative $\tilde{\nabla}_{k}\hat{\Fcal}_{k0}$ of the non-Abelian electric field is exactly the right hand side of the Yang-Mills electrostatic equation. The general gauge symmetry arguments of Ref.~\cite{
Tokatly2008} (used there to uncover the nature of the equilibrium spin currents) show that $\tilde{\nabla}_{k}\hat{\Fcal}_{k0}$ is proportional of the magnetization induced in the F-region by a non-uniform exchange field and/or SO coupling. This reveals the nature of the second term in Eq.~(\ref{covUsadel-tperp}) and provides an interesting connection between the generation of LRTC at the edges of Neel domain walls, and the Yang-Mills electrostatics.

We now analyze a SO-generated LRTC in the case of a uniform magnetization. A special type of SO coupling in Eq.~(\ref{H-linearSO}) should naturally occur in the vicinity of hetero-interfaces {where inversion asymmetry exists}\cite{Rashba,Edelstein,Takei}. Hence we concentrate on the situation when the SU(2) vector potential is localized around the S/F interface and has  in-plane components $\hat{\Acal}_x$ and $\hat{\Acal}_y$. This implies that only $\hat{\Fcal}_{x0}$ and $\hat{\Fcal}_{y0}$ are nonzero. 

As a first example we consider a transversal S/F structure (Fig.~1a) assuming for definiteness a Rashba-Dresselhaus SO term with $\hat{\Acal}_x=\beta\hat\sigma_x-\alpha\hat\sigma_y$ and $\hat{\Acal}_y=(\alpha\hat\sigma_x-\beta\hat\sigma_y)$, where $\alpha$ and $\beta$ are the Rashba and Dresselhaus constants.  We assume a contstant in-plane magnetization. Thus,  only the second term in the r.h.s of  Eq.~(\ref{covUsadel-tperp}) serves as source for the  LRTC. One can easily show that  this term is non-zero only if $\alpha\beta\neq 0$ and ${\cal A}_0^x\neq{\cal A}_0^y$. This in particular means that a pure Rashba or Dresselhaus SO coupling does not induce the LRTC in a transversal geometry with an in-plane magnetization.   { We emphasize that the LRTC discussed here has  s-wave symmetry, in contrast with the odd in momentum  triplet component predicted in Refs.\cite{Edelstein,Takei} for  pure ballistic S/F and S/N systems  in the presence of an interface SO coupling. }  

Lateral S/F structures are more favorable for the existence of LRTC. Consider the structure  shown in  {Fig.~1b}. Assuming for simplicity, but without loss of generality, that  the F film is thin enough, we integrate Eqs.~(\ref{covUsadel-s}), (\ref{covUsadel-tpar})-(\ref{covBC-tperp}) over $z$-direction, and obtain the following set of 1D equations
\begin{eqnarray}
 \label{Usadel-lats}
&&\partial_{x}^2f_s - \kappa_\omega^2 f_s -2i\frac{\sgn\omega}{D} |\hat{\Acal}_0|^2f_t^{\parallel}=-\theta(-x)\bar\gamma f_{\Delta},\\
\label{Usadel-latpar}
&&\partial_{x}^2f_t^{\parallel} - \kappa_\omega^2f_t^{\parallel} -\frac{2i\sgn\omega}{D} f_s=0,\\
\label{Usadel-latperp}
&&\partial_{x}^2\hat{\tf}_t^{\perp} - \kappa_\omega^2\hat{\tf}_t^{\perp} =
- 2\theta(-x)\hat{\bar{\Fcal}}_{x0}\partial_{x}f_t^{\parallel},
\end{eqnarray}
where  $\bar\gamma$ and ${\bar{\Fcal}}_{x0}$ are effective values averaged over the thickness.
The boundary conditions at $x=0$ are the continuity of all functions, and the continuity of $\partial_xf_s$, $\partial_xf_t^{\parallel}$, and $\partial_{x}\hat{\tf}_t^{\perp} + \hat{\Fcal}_{x0}f_t^{\parallel}$. This boundary problem can be solved straightforwardly. 
In the interesting limiting case of the exchange field $|\hat{\Acal}_0|$ much larger than the superconducting gap $\Delta$ the condensate function at $x>0$ ({\it cf.} {Fig. 1b}) is given by the expression
\begin{equation}
\label{decay}
\hat{\tf}(x) = \sum_{\alpha}C_{\alpha}e^{-\kappa_{\alpha} x}\left[1+(-1)^{\alpha}{\hat{\Acal}_0}/{|\hat{\Acal}_0|}\right]+ \hat{\bar{\Fcal}}_{x0}^{\perp} Ce^{-\kappa_\omega x}\; ,
\end{equation}
where $\alpha=1,2$, $C_{1,2}\approx-\bar\gamma f_{\Delta}D^2\kappa_{2,1}^2/(4|{\cal A}_0|^2)$, $\kappa_{1,2}^2={\pm2i\sgn\omega|{\cal A}_0|/D}$, with ${\rm Re}\kappa_{1,2}>0$ and 
$C\approx-(3i/2)\bar\gamma f_{\Delta}D\sgn\omega/(|{\cal A}_0|^2\kappa_\omega)$.
All components decay away from the edge plane $x=0$. The singlet and parallel to $\hat{\cal A}_0$ triplet components [the first term in Eq.~(\ref{decay})] decay over the short magnetic distance $\sqrt{D/2h}$, while  the triplet component perpendicular to $\hat{\cal A}_0$ [the second term in Eq.~(\ref{decay})] decays over a larger length of the order of $\sqrt{D/2T}$ confirming its long-range character. {Figure~1b} shows tha the LRTC decays in both directions from the inhomogeneity at $x=0$, which looks very similar to the LRTC generated at the edge of a Neel domain wall \cite{Fominov2005}. This similarity is not accidental. In fact, in the particular case of $\hat{\Acal}_y=0$ our system is gauge-equivalent to the Neel wall with an edge at $x=0$.

We now consider a lateral structure with two S-electrodes separated by a distance $L$ (see {Fig.~1c}). If $L\gg\sqrt{D/2h}$ the Josephson coupling is only mediated by the LRTC, and the critical current is given by 
\begin{equation}
I_c=(\frac{{\cal S}\sigma_F}{e})\tr(\hat{\bar{\Fcal}}_{x0}^{\perp})^2T\sum_{\omega_n}\kappa_\omega C^2(\omega_n)e^{-\kappa_\omega L}\; ,\label{current}
\end{equation} 
where ${\cal S}$ and $\sigma_F$ are the cross section and conductivity of the F-region.
From this equation one concludes that a finite SU(2) electric field with a component perpendicular to the magnetization is the source of the long-range Josephson effect. 
The lateral geometry shown in {Fig.~1c} is equivalent to the one explored in the experiments of Refs.\cite{Keizer2006,Wang2010}. Thus, our theory gives a plausible explanation for the long-range effects observed 
in these experiments. One argues in that  case that the long-range Josephson current is either due to a SO coupling at the S/F interfaces \cite{Keizer2006}, or to a Rashba-type SO coupling in  quasi-1D geometry of  Ref. \cite{Wang2010}.   A triplet component can also be induced in a SNS lateral structure with Rashba SO coupling in an external Zeeman field\cite{Malshukov}.  
\begin{figure}[t]
\begin{center}
\includegraphics[width=\columnwidth]{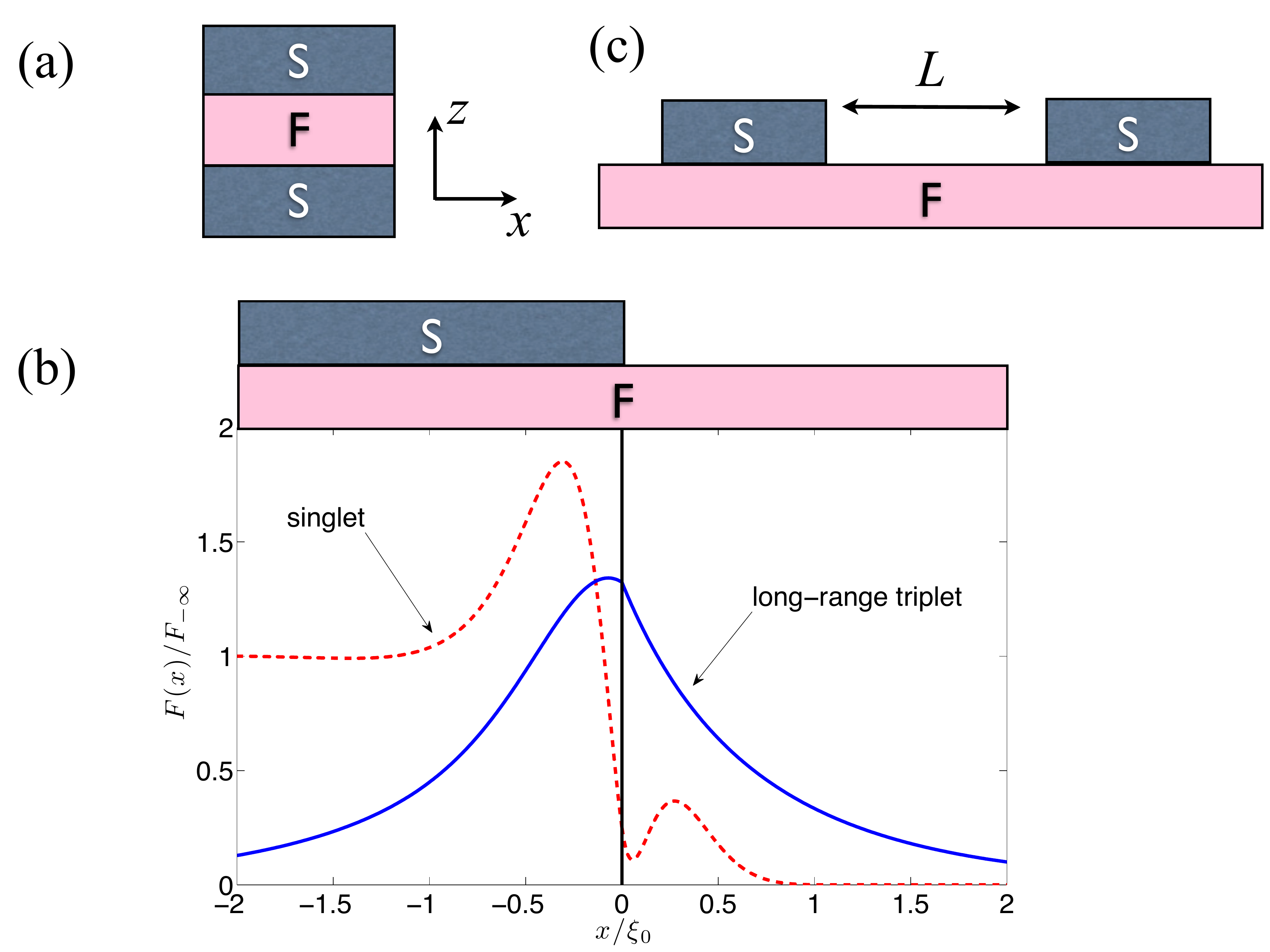}
\caption{Different geometers considered in the text. (a) The transversal S/F/S junction. (b) The lateral S-F structure and space dependence of $F=\sum_\omega|f|^2$ for the singlet and LRTC normalized with the asymptotic value at $x=-\infty$. Here $\xi_0=\sqrt{D/\Delta}$, $h=10\Delta$,  $T=0.05\Delta$ and $\Delta$ is the order parameter in the superconductor. (c) The lateral Josephson junction}
\end{center}
\end{figure}

We finally generalize our results to ferromagnets with a generic momentum dependent effective exchange field. Our starting point is the following Hamiltonian in the F-region
\begin{equation}
H= \xi_\pv-[b^a(\pv)+h^a(\pv)]\sigma^a\;.\label{HF}
\end{equation} 
Here $\xi({\pv})$ is the spin-independent part of the quasiparticle energy. The spin-dependent contribution is written as the sum of an even, $\hat{h}({-\pv})=\hat{h}({\pv})$, and  an odd, $\hat{b}(-{\pv})=-\hat{b}({\pv})$, in momentum parts.  $\hat{h}({\pv})$
 describes the Zeeman-type exchange term, which is, in general, momentum dependent in realistic systems. The odd part $\hat{b}({\pv})$ corresponds to a generic SO interaction which preserves the time reversal symmetry{\cite{noteH}  }

{Following the standard procedure} (see for example Ref.\cite{LObook}), one derives the Eilenberger equation \cite{Eilenberger} for a generic spin-dependent Hamiltonian
\begin{equation}
v^{F}_{k}(\nv)\partial_{k} \check g+[\omega\tau_3,\check g]-i[\hat{h}({\bf n})\tau_3+\hat{b}({\bf n}),\check g]=-\frac{1}{2\tau}[\langle\check g\rangle,\check g]\label{Eil}
\end{equation}
where  ${\nv}$ is a unit vector pointing in the direction of momentum, $v^{F}_{k}(\nv)$ are components of the Fermi velocity, $\tau$ is the impurity scattering time, and $\langle ...\rangle$ denotes averaging over  ${\nv}$. This equation allows for a general anisotropy with different velocities and spin splittings at different points on the Fermi surface. {In the particular case of an isotropic $h$, Rashba SO-coupling and pure ballistic system we recover the equation used in Ref.~\cite{BuzdinSO}.}
We focus here  on the diffusive limit, in which $\tau^{-1}$ determines the largest energy scale in Eq.~(\ref{Eil}). In this case Eq.~(\ref{Eil}) reduces to the Usadel equation for the angle-averaged GF \cite{Usadel,noteV}.
For a general anisotropic ferromagnet the linearized Usadel equation  takes the form
\begin{eqnarray}
&D_{kj}\partial_{k}\partial_{j}\hat f - 2\omega\hat f& 
-{i\sgn\omega}\left\{ \langle\hat{h}\rangle,\hat f\right\}-
2i\tau\left[\langle v^{F}_{k}\hat{b}\rangle,\partial_k\hat f\right]- \label{linUS}\\
&-i\tau\left[\partial_k\langle v^{F}_{k}\hat{b}\rangle,\hat f\right]&
-\tau\langle\left[\hat b,\left[\hat b, \hat f\right]\right]\rangle - 
\tau\langle\left\{\delta\hat h,\left\{\delta\hat h,\hat f\right\}\right\}\rangle=0\nonumber\; .
\end{eqnarray}
 where  $D_{kj}=\tau\langle v^{F}_{k}v^{F}_{j}\rangle$ is the tensor of diffusion coefficients, and $\delta\hat{h}({\nv})=\hat{h}({\nv})- \langle\hat{h}\rangle$ is the variation of the Zeeman field at the Fermi surface. 
With the exception of the last term containing $\delta \hat h$,  Eqs.~(\ref{linUS}) and the gauge covariant Usadel equation,  Eq.~(\ref{covUsadel-gen}), are structurally equivalent. The first three terms in Eq.~(\ref{linUS}) correspond to  Eq.~(\ref{covUsadel-gen}) without the commutators coming from the covariant gradients.  These terms in Eq.~(\ref{covUsadel-gen}) correspond to the commutators in Eq.~(\ref{linUS}) involving the SO field $\hat b$. Because of this similarity our analysis of Eq.~(\ref{covUsadel-gen}) is directly applicable to Eq.~(\ref{linUS}). Hence we conclude that the last four terms in Eq.~(\ref{linUS}) serve as a the source for the LRTC. More precisely, 
the LRTC is generated if any of these terms has a finite component perpendicular to the exchange field $\langle\hat{h}\rangle$ averaged over the Fermi surface. We can draw a  remarkable  conclusion from this result:  From the knowledge of the electronic properties at the Fermi level of S/F systems, namely, from $\xi({\pv})$, $\hat h(\pv)$, and $\hat b(\pv)$ in Eq.~(\ref{HF}) one can easily infer whether or not the LRTC would exist in the hybrid structure. Moreover, Eq.~(\ref{linUS}) is quite general and can be used in a  broad context of problems involving superconductivity and spin-fields.

In conclusion, we  presented a general description of the long-range triplet superconductivity in S/F structures. Starting from the linear in momentum SO coupling we  developed the SU(2) covariant theory describing  the diffusion of the condensate and identified the SU(2) electric field as a physical source for the LRTC. We  also considered the case of an arbitrary momentum dependence of the spin-fields and derived a useful equation from which, by knowledge of the electronic structure of the ferromagnet and the interfaces, one can directly predict whether the LRTC is generated or not.  Our results not only unify in an elegant way all models describing the long-range proximity effect in S/F structures, but also predict new sources for  the singlet-triplet conversion and provide a useful tool in the search for triplet superconducting  correlations.

{\it Acknowledgements} We thank  Ivo Souza for useful discussions. F.S.B thanks Martin Holthaus and his group for their kind
hospitality at the Physics Institute of the Oldenburg University. The work of F.S.B  was supported by  the Spanish Ministry of Economy and Competitiveness under
Project FIS2011-28851-C02-02 and the Basque Government under UPV/EHU Project IT-366-07.  {I.V.T.  acknowledges funding by the ``Grupos Consolidados UPV/EHU del
Gobierno Vasco'' (IT-319-07) and Spanish MICINN (FIS2010-21282-C02-01).}


\begin{thebibliography}{10}
\bibitem{noteSO} {The SO-coupling entering Hamiltonian (\ref{H-linearSO}) originates from the electronic band structure and/or geometric constraints. This should be contrasted to the SO effect of random impurities,  studied in Ref.~\cite{spinorbit}. While in the latter case no triplet component is created, the SO term in Eq.~(\ref{H-linearSO}) may generate the LRTC. }
\bibitem{Note}{Formally the Wilson link operator  is defined via the following path-ordered exponential $\hat{W}(\rv_2,\rv_1)=P\exp{[i\int_{C_{12}}\hat{\Acal}_jdx_j]}$, where the integration path $C_{12}$ goes from $\rv_1$ to $\rv_2$ along the straight line \cite{Heinz}.}
\bibitem{noteH}{ Note that Hamiltonian (\ref{H-linearSO}) is a particular case of Hamiltonian (\ref{HF}) for a momentum independent $h$ field and a linear in momentum $b$.}
\end{thebibliography}
\end{document}